\def\CMP{\sevenrm Commun.\ Math.\ Phys}
\def\JMP{\sevenrm J.\ Math.\ Phys}
%
%
\def\today{\number\day .\space\ifcase\month\or
January\or February\or March\or April\or May\or June\or
July\or August\or September\or October\or November\or December\fi, \number \year}
%
%
\newcount \theoremnumber
\def\cleartheoremnumber{\theoremnumber = 0 \relax}

\def\Prop #1 {
             \advance \theoremnumber by 1
             \vskip .6cm 
             \goodbreak 
             \noindent
             {\bf Propsition {\the\headlinenumber}.{\the\theoremnumber}.}
             {\sl #1}  \goodbreak \vskip.8cm}

\def\Th#1 {
             \advance \theoremnumber by 1
             \vskip .6cm  
             \goodbreak 
             \noindent
             {\bf Theorem {\the\headlinenumber}.{\the\theoremnumber}.}
             {\sl #1}  \goodbreak \vskip.8cm}

\def\Lm#1 {
             \advance \theoremnumber by 1
             \vskip .6cm  
             \goodbreak 
             \noindent
             {\bf Lemma {\the\headlinenumber}.{\the\theoremnumber}.}
             {\sl #1}  \goodbreak \vskip.8cm}

\def\Cor#1 {
             \advance \theoremnumber by 1
             \vskip .6cm  
             \goodbreak 
             \noindent
             {\bf Corollary {\the\headlinenumber}.{\the\theoremnumber}.}
             {\sl #1}  \goodbreak \vskip.8cm} 
%
%
\newcount \equationnumber

\newcount \refnumber

\def\[]    {\global 
            \advance \refnumber by 1
            [{\the\refnumber}]}

\def\# #1  {\global 
            \advance \equationnumber by 1
            $$ #1 \eqno ({\the\equationnumber}) $$ }

\def\% #1 { \global
            \advance \equationnumber by 1
            $$ \displaylines{ #1 \hfill \llap ({\the\equationnumber}) \cr}$$} 

\def\& #1 { \global
            \advance \equationnumber by 1
            $$ \eqalignno{ #1 & ({\the\equationnumber}) \cr}$$}
%
%
%
\def\REF #1 #2 #3 #4 #5 #6 #7   {{\sevenbf [#1]}  & \hskip -9.5cm \vtop {
                                {\sevenrm #2,} 
                                {\sevensl #3,} 
                                {\sevenrm #4.} 
                                {\sevenbf #5,} 
                                {\sevenrm #6,} 
                                {\sevenrm (#7)}}\cr}
\def\BOOK #1 #2 #3 #4  #5   {{\sevenbf [#1]}  & \hskip -9.5cm \vtop {
                             {\sevenrm #2,}
                             {\sevensl #3,} 
                             {\sevenrm #4,} 
                             {\sevenrm #5.}}\cr}
%
%
\def\bull{$\sqcup \kern -0.645em \sqcap$}
%
%

\def\Rem#1{\vskip .4cm \goodbreak \noindent
                                     {\it Remark.} #1 \goodbreak \vskip.5cm }

\def\Pr#1{\goodbreak \noindent {\it Proof.} #1 \hfill \bull  \goodbreak \vskip.5cm}
%
%
\newcount \headlinenumber

\newcount \headlinesubnumber
\def\clearheadlinesubnumber{\headlinesubnumber = 0 \relax}
\def\Hl #1 {\goodbreak
            \cleartheoremnumber
            \clearheadlinesubnumber
            \advance \headlinenumber by 1
            {\bf \noindent {\the\headlinenumber}. #1}
            \nobreak \vskip.4cm \rm \noindent}

\font\fourteenrm=cmr10 at 14pt
\font\sevensl=cmsl10 at 7pt
\font\sevenit=cmti7 at 7pt

\font\css=cmss10
\font\Rosch=cmr10 at 9.85pt
\font\Cosch=cmss12 at 9.5pt
\font\rosch=cmr10 at 7.00pt
\font\cosch=cmss12 at 7.00pt
\font\nosch=cmr10 at 7.00pt
%
%
%
%
%
\def\Z                 {\hbox{{\css Z}  \kern -1.1em {\css Z} \kern -.2em }}
\def\R                 {\hbox{\raise .03ex \hbox{\Rosch I} \kern -.55em {\rm R}}}
\def\N                 {\hbox{\rm I \kern -.55em N}}
\def\C                 {\hbox{\kern .20em \raise .03ex \hbox{\Cosch I} \kern -.80em {\rm C}}}

\def\r                 {\hbox{\raise .03ex \hbox{\rosch I} \kern -.45em \hbox{\rosch R}}}
\def\n                 {\hbox{\hbox{\rosch I} \kern -.45em \hbox{\nosch N}}}
\def\c                 {\hbox{\raise .03ex \hbox{\cosch I} \kern -.70em \hbox{\rosch C}}}

\def\z                 {\hbox{\kern 0.2em {\cal z}  \kern -0.6em {\cal z} \kern -0.3em  }}
\def\1                 {\hbox{\rm \thinspace \thinspace \thinspace \thinspace
                                  \kern -.50em  l \kern -.85em 1}}
\def\unit                 {\hbox{\sevenrm \thinspace \thinspace \thinspace \thinspace
                                  \kern -.50em  l \kern -.85em 1}}
%
%

\def\A                 {{\cal A}}

\def\M                 {{\cal M}}
\def\H                 {{\cal H}} 
 
\def\O                 {{\cal O}}

\nopagenumbers
\def\Draft  {\hbox{Preprint \today}}
\def\firstheadline{\hss \hfill  \Draft  \hss} 
\headline={
\ifnum\pageno=1 \firstheadline
\else 
\ifodd\pageno \rightheadline 
\else \leftheadline \fi \fi}

\def\rightheadline{\sevenrm DECAY OF SPATIAL CORRELATIONS IN THERMAL STATES
\hfill \folio } 
\def\leftheadline{\sevenrm \folio \hfill CHRISTIAN D.\ J\"AKEL}
\voffset=2\baselineskip
\magnification=1200
%
%
%
%
%

\vskip 1cm
                  
\centerline{\fourteenrm Decay of Spatial Correlations}
                  
\vskip .2cm
\centerline{\fourteenrm in Thermal States}
                  
\vskip .8cm
                  
\noindent
CHRISTIAN D.\ J\"AKEL\footnote{$^\star$}{\sevenrm 
present address: Inst.\ f.\ Theoretische Physik, Universit\"at Wien, 
Boltzmanng.\ 5, A-1090 Wien, Austria, e-mail: cjaekel@nelly.mat.univie.ac.at} 

\noindent
{\sevenit II. Institut f\"ur Theoretische Physik,
Universit\"at Hamburg, D-22761 Hamburg,
Federal Republic of Germany}

\vskip .4cm     
\noindent {\sevenbf Abstract}. {\sevenrm 
We study the cluster properties of thermal equilibrium states in theories with a maximal
propagation velocity (such as relativistic QFT). Our analysis, carried out in the setting of
algebraic quantum field theory, shows that there is a tight relation 
between spectral properties of the generator of time translations 
and the decay of spatial
correlations in thermal equilibrium states, in complete analogy to the 
well understood case of the vacuum state.} 


\vskip .8 cm

\noindent
D\'ecroissance des corr\'elations spatiales pour les \'etats thermaux.

\vskip .4cm     
\noindent {\sevenbf R\'esum\'e.} {\sevenrm 
Nous \'etudions la propri\'et\'e de cluster des \'etats
thermaux d'\'equilibre dans des th\'eories poss\'edant une vitesse
maximale de propagation (comme les th\'eories des champs
relativistes).
Notre analyse, men\'ee dans le cadre de la th\'eorie alg\'ebrique
des champs, montre qu'il existe une relation \'etroite entre les
propri\'et\'es spectrales du g\'en\'erateur des translations
temporelles et la d\'ecroissance des corr\'elations spatiales pour les
\'etats thermaux d\'equilibre, en compl\'ete analogie avec le cas bien
connu de l'\'etat de vide.}

%
%
%
%
%

\vskip 1cm

\Hl{Introduction}

\noindent
The cluster theorem of relativistic quantum field theory states 
that correlations of
local observables in the vacuum state decrease at least like $\delta^{-2}$ 
if $\delta$ is their space-like separation~[AHR]. In the presence of a mass-gap
one has even an exponential decay like~${\rm e}^{-M \delta}$, where~$M$ is the minimal
mass in the theory~[AHR],~[F].

Little is known about the decay of correlations in the case of thermal equilibrium states.
But a few remarks are in order: a) The spectrum of the generators of translations 
is all of $\R^4$, i.e., one cannot distinguish 
between theories with short respectively long range forces by looking at the shape of 
the spectrum. 
Yet, as we shall see, spectral properties of the generators are still of importance. 
b) The decay of correlations may be weaker than in the vacuum case. Simple examples are KMS states
in the theory of free massless bosons, where correlations between space-like separated observables 
decay only like $\delta^{-1}$. 
c)~Due to the KMS condition, which is characterizing for
equilibrium states, the main contribution to the spatial correlations 
comes from low energy excitations; high energy excitations are suppressed in an
equilibrium state. The KMS condition paves the way for a model independent analysis: If we 
have some information on
the spectral properties of the generator ${\bf H}_\beta$ of the time evolution in the
thermal representation,
then we can specify the decay of spatial correlation functions. 

We start with a list of assumptions and
properties relevant for the thermal states of a local quantum field theory. We refer
to [H], [HK] for further discussion and motivation. 

\vskip .5cm
\noindent
(i)   {\sl (Local observables)}. The central object in the mathematical description
of a theory is an inclusion preserving map 
\# {\O \to \A (\O),}
which assigns to any open bounded region $\O$ in Minkowski space $\R^4$ a unital
$C^*$-algebra.
The Hermitian elements of $\A (\O)$ are interpreted as the observables which
can be measured at times and locations in $\O$.
Thus the net is {\sl isotonous}, 
\# { \O_1 \subset \O_2 \Rightarrow \A(\O_1) \subset \A(\O_2).}
Isotony allows one to embed $\A(\O)$ in the  {\sl algebra of quasilocal observables} 
\# { \A = 
\overline{ \bigcup_{ {\cal O}  \subset \r^4} \A(\O) }^{C^*}  .} 

\vskip .2cm
\noindent
(ii)  {\sl (Einstein-Causality {\it or} Locality)}. The net $\O \to \A (\O)$ is local: 
Let $\O_1$ and $\O_2$ denote two arbitrary space-like separated open, 
bounded regions. Then 
\# {\A (\O_1) \subset \A(\O_2)' \quad \hbox{\rm for} \quad \O_1 \subset \O_2',}
where $\O'$ denotes the space-like complement of $\O$. Here $\A(\O)'$ denotes the set of operators 
in~$\A$ which commute with all operators in $\A(\O)$.
Thus
\# { [ a, b ] := ab - ba =  0  } 
for all  $a \in \A(\O_1)$, $b \in \A(\O_2)$.

\vskip .2cm
\noindent
(iii)  {\sl (Time-translations)}. 
The time evolution acts as a strongly continuous group of automorphisms on $\A$,
and it respects the local structure, i.e.,
\# {  \tau_t (\A (\O)) 
= \A (\O +te) ,} 
for all $ t \in \R$. Here $e$ is a unit vector denoting the time 
direction with respect to a given Lorentz-frame. 
 
\vskip .5cm

\noindent
We now recall how equilibrium states are characterised within the set of states
[BR], [E], [S], [T], [R]. Heuristically, one expects an equilibrium state to have
the following properties: 
a)~Time-evolution invariance. b) Stability
against small adiabatic perturbations.
Together with a few (physically motivated) technical assumptions, these properties
lead [HKTP] to the subsequent
criterion [HHW], named after Kubo [Ku], Martin and Schwinger [MS]\footnote{$^\dagger$}
{\sevenrm Recently it was shown in [BB] that in a relativistic theory
the thermal correlation functions have stronger analyticity 
properties in configuration space than those imposed by the KMS condition. These analyticity
properties may be understood as a remnant of the relativistic spectrum condition in the
vacuum sector and lead to a Lorentz-covariant formulation of the KMS-condition.}. 

\vskip .5cm
\noindent
(iv) {\sl (KMS-condition)}.  A state
$\omega_\beta $ satisfies the KMS-condition at inverse temperature $\beta > 0$ iff for
every pair of elements $a, b \in \A$ there exists an analytic function~$F_{a,b}$ in the strip 
$S_\beta = \{ z \in \C | 0 < \Im z < \beta \}$ with continuous boundary values at $\Im z =0$
and $\Im z = \beta$, given by, 
\# {F_{a,b} (t) = \omega_\beta ( b \tau_t ( a)), 
\quad F_{a,b} (t + i \beta ) = \omega_\beta (\tau_t ( a) b ), \qquad
\forall t \in \R.}

\vskip .5cm

We recall that given a KMS state $\omega_\beta $, the GNS-construction provides a 
representation~$\pi_\beta$ of $\A$ on a Hilbert space $\H_\beta$, together with a unit vector
such that
$\omega_\beta  (a) = (\Omega_\beta, \pi_\beta (a) \Omega_\beta)$, for all $ a \in \A$.  
Moreover, $\Omega_\beta$ is
cyclic and separating  for $\pi_\beta(\A)''$.
If $\omega_\beta$ is locally normal w.r.t.\ the vacuum, 
then the KMS condition implies that  
$\Omega_\beta$ is cyclic even 
for the strictly local algebras $ \pi_\beta(\A  (\O ))$, i.e.,
\# { \H_\beta = \overline { \pi_\beta(\A  (\O )) \Omega_\beta },}
for arbitrary open space-time regions $ \O  \in \R^4$ [J]. 
The function $F_{ 1 , b}(z)$, $z \in  {\cal S}_\beta $ satisfies 
$F_{ 1 , b}(t) = F_{ 1 , b} (t + i \beta)$. Thus  $F_{ 1 , b}(t)$ extends to a  
periodic, bounded entire function. By 
Liouville's theorem $F_{ 1 , b}(z)$ is constant, $\omega_\beta  \circ \tau_t =
\omega_\beta $. Thus
\# { U  (t) \pi_\beta (a) \Omega_\beta 
= \pi_\beta ( \tau_t (a))  \Omega_\beta ,
\qquad \forall
a \in \A  ,}
defines a strongly continuous one-parameter group of unitary operators 
$\{ {\bf U} (t) \}_{t \in \r}$
implementing the time-evolution. 
By Stone's theorem there exists a self-adjoint generator ${\bf H}_\beta$
such that 
${\bf U} (t) = {\rm e}^{i {\bf H}_\beta  t} $ for all $t \in \R$.
For $0 \le \beta < \infty$ the operator ${\bf H}_\beta$
is not bounded from below, its spectrum is symmetric and
consists typically of the whole real line. 
Restricting attention to pure phases we assume that  
$\Omega_\beta$ is the unique --- up to a phase
--- time-invariant vector in~$\H_\beta$.

\vskip 1cm

\Hl{Analytic Properties of Thermal Correlation Functions}

\noindent
An equilibrium state distinguishes a rest frame~[N] [O] and thereby 
destroys relativistic covariance.
But the key feature of a relativistic theory,
known as  ``Nahwirkungsprinzip'' or locality, survives: 
Given two regions $\O_1$ and $\O_2$,
separated by a space-like 
distance $\delta > 0$ such that $\O_1 + t e \subset \O_2$, $|t| < \delta$,
the commutator of two local obsevables 
$a \in \A(\O_1)$, $b \in \A(\O_2)'$  vanishes for $|t| < \delta$, i.e., 
\# { [a,\tau_t (b)] = 0,}
for $|t| < \delta$. Thus locality together with the KMS condition
implies that the function
\# {F_{a,b} \colon t \to
\omega_\beta (b \tau_t (a)) }
can be analytically continued into the infinitely often cut plane  
\# {  {\cal I}_\delta = \C \backslash \{ z \in \C | 
\Im z = k \beta, k \in \Z, |\Re z| \ge \delta \}. }
Set ${\bf A} = \pi_\beta (a) $, 
${\bf B} \in \pi_\beta (b) $, then  $F_{a,b} \colon  {\cal I}_\delta \to \C$ is given by
\# {
F_{a,b} (t + i \alpha) 
:= 
\left\{
\eqalign{
&  { (\Omega_\beta , {\bf B} {\rm e}^{ i ( t + i \alpha - i \beta ) 
{\bf H}_\beta} {\bf A} \Omega_\beta ), }
\cr
&
{ (\Omega_\beta , {\bf B} {\rm e}^{ i ( t + i \alpha ) 
{\bf H}_\beta} {\bf A} \Omega_\beta ), } 
}
\right\} 
{\rm \ for \ }  
\left\{
\eqalign{
& { \beta  < \alpha < 2\beta,}  
\cr
& {0 \le \alpha \le  \beta ,}  }  
\right\} 
}
and periodic extension in $\Im z$ with period $2\beta$. 

\goodbreak

The periodic structure of the cuts can now be explored\footnote{$^\dagger$}
{\sevenrm The author acknowledges stimulating discussions with J. Bros on this 
point, the present solution is due to D. Buchholz.}.
Let us consider the spectral projections ${\bf P}^+$, ${\bf P}^-$ and ${\bf P}_\beta =
| \Omega_ \beta) ( \Omega_\beta |$  
onto the strictly positive, the strictly negative, and the discrete 
spectrum  $\{ 0 \} $ of ${\bf H}_\beta$.
The function $f_+ \colon \{ z \in \C | \Im z \ge 0 \} \to \C$,
\# { f_+(z) =
(\Omega_\beta \, , \, {\bf B}  {\rm e}^{ iz {\bf H}_\beta} {\bf P}^+ {\bf A} \Omega_\beta)
- (\Omega_\beta \, , \, {\bf A} {\rm e}^{- iz {\bf H}_\beta} {\bf P}^-  {\bf B} \Omega_\beta) }
is analytic in the upper half plane $\Im z > 0$, and continuous for $\Im z \searrow 0$,
the function 
$f_- \colon \{ z \in \C | \Im z \le 0 \}  \to \C$,
\# { f_-(z) =
- (\Omega_\beta \, , \, {\bf B} {\rm e}^{ iz {\bf H}_\beta} {\bf P}^- {\bf A} \Omega_\beta) 
+ (\Omega_\beta \, , \, {\bf A} {\rm e}^{ -iz {\bf H}_\beta} {\bf P}^+ {\bf B} \Omega_\beta)}
is analytic in the lower half plane $\Im z < 0$, and continuous for $\Im z \nearrow 0$.
The discrete spectral value $\{ 0 \}$ gives no contribution, $f_{ \scriptstyle \{ 0 \} } \colon
\C \to \C$, 
\& { f_{ \scriptstyle \{ 0 \} } (z) &=
(\Omega_\beta \, , \, {\bf B}  {\rm e}^{ iz {\bf H}_\beta} \Omega_\beta)(\Omega_\beta \, , \,
{\bf A} \Omega_\beta)
- (\Omega_\beta \, , \, {\bf A} {\rm e}^{- iz {\bf H}_\beta} \Omega_\beta)
(\Omega_\beta \, , \, {\bf B} \Omega_\beta) 
\cr
&= 0 , \qquad \forall z \in \C. }
Thus we can decompose the commutator $\omega_\beta  ( [ b  \, , \, \tau_t (a) ]) $
into two pieces:
\# {   (\Omega_\beta , \pi_\beta ( [ b  \, , \, \tau_t (a) ]) \Omega_\beta) = 
f_+ (t) - f_- (t)  
\qquad \forall t \in \R .}
The l.h.s.\ vanishes for $|t| < \delta$, i.e., 
the boundary values of the function defined in (14) and (15) (from the upper and 
the lower half plane, respectively)
coincide for $| \Re z| < \delta $.
Using the Edge-of-the-Wedge Theorem [SW] one concludes that there is a function~$f_{a,b}$ which
is analytic on the twofold cut plane
$  {\cal P}_\delta = \C \backslash \{ z \in \C | \Im z = 0,  |\Re z | \ge \delta \} $
such that
\# { f_{a,b} (z) = 
\left\{
\hskip -.1cm
\eqalign{
&  {  (\Omega_\beta  ,  {\bf B} {\rm e}^{ iz {\bf H}_\beta} {\bf P}^+ {\bf A} \Omega_\beta)
- (\Omega_\beta  ,  {\bf A} {\rm e}^{- iz {\bf H}_\beta} {\bf P}^-  {\bf B} \Omega_\beta) }
\cr
&  {  (\Omega_\beta  , {\bf A} {\rm e}^{ -iz {\bf H}_\beta} {\bf P}^+ {\bf B} \Omega_\beta)
- (\Omega_\beta  ,  {\bf B} {\rm e}^{ iz {\bf H}_\beta} {\bf P}^- {\bf A} \Omega_\beta) }
}
\hskip -.1cm
\right\} 
{\rm  for }  
\left\{
\hskip -.1cm
\eqalign{
& {  \qquad \Im z > 0,}  \cr
& { \left\{ \eqalign{ & { \Im z = 0, | \Re z | < \delta ,}  \cr
                      & {  \Im z < 0 .}                                  } \right\} } 
} 
\hskip -.1cm
\right\} 
}
Since ${\bf P}^{+} \notin  \pi_\beta (\A)$ the KMS condition does not apply.
But we can approximate ${\bf P}^{+}{\bf A} \Omega_\beta$ with the help of analytic elements 
${\bf A}_\epsilon \in \pi_\beta (\A_\tau)$ 
such that\footnote{$^\sharp$}{\sevenrm Here $\scriptstyle {\cal A}_\tau$ denotes the set of entire 
analytic elements for $\scriptstyle \tau$ [BR, 2.5.20]. Note that $\scriptstyle {\cal A}_\tau$
is a norm dense *-subalgebra of $\scriptstyle {\cal A}$ [BR, 2.5.22].}
\# { \lim_{\epsilon \to 0} {\bf A}_\epsilon \Omega_\beta
= {\bf P}^{+} {\bf A} \Omega_\beta,
\qquad  \lim_{\epsilon \to 0}  {\bf A}^*_\epsilon \Omega_\beta
= {\bf P}^{-} {\bf A}^* \Omega_\beta.}
Set
\# { {\bf A}_\epsilon 
= {1 \over \sqrt {2 \pi}} 
\int {\rm d}t \,  g_\epsilon (t)   \, {\rm e}^{ i {\bf H}_\beta t} {\bf A}
{\rm e}^{- i {\bf H}_\beta t} , \qquad \forall {\bf A} \in \pi_\beta(\A)  ,}
where $ g_\epsilon \in S(\R)$ are complex test functions with uniformly bounded Fourier
transform and
\# {
\tilde g_\epsilon (\nu) = 
\left\{
\eqalign{
& 0
\cr
& 1
\cr
& 0}
\right\} 
{\rm \ for \ }  
\left\{
\eqalign{
& {  \nu \in (- \infty , \epsilon / 2] ,}  
\cr
& { \nu \in [\epsilon , 1/ \epsilon]}  
\cr
& { \nu \in [\epsilon + 1/ \epsilon, \infty).}  }  
\right\} 
}

\goodbreak

\noindent
The Bochner integral (20) exists\footnote{$^\star$}{\sevenrm
Since $\scriptstyle \tau_t$ is strongly continuous and $\scriptstyle g_\epsilon \in S(\r)$,
there exists a sequence of countably valued functions
$\scriptstyle t \to a_n (t)$, $\scriptstyle n \in \n$,
converging almost everywhere to $\scriptstyle t \to g_\epsilon (t) \tau_t (a)$.}
for all
$\epsilon > 0 $ and ${\bf A}_\epsilon \in \pi_\beta (\A_\tau)$.
The spectral resolution of ${\bf H}_\beta$ implies
\# { {\bf A}_\epsilon \Omega_\beta = {1 \over \sqrt {2 \pi}} \int {\rm d}t
\,  g_\epsilon (t)   \, {\rm e}^{i {\bf H}_\beta t} {\bf A} \Omega_\beta
= \tilde g_\epsilon ({\bf H}_\beta) {\bf A} \Omega_\beta .}
The sequence $\tilde g_\epsilon$ converges uniformly on compact
sets in $\R \backslash \{ 0 \}$ to the Heaviside step-function. 
By assumption, the Fourier transforms $\tilde g_\epsilon$ are uniformly bounded.
Thus the spectral theorem 
yields $\lim_{\epsilon \to 0} \| \tilde g_\epsilon ({\bf H}_\beta) {\bf A} \Omega_\beta
- {\bf P}^+ {\bf A} \Omega_\beta \| = 0$,  for all ${\bf A} \in \pi_\beta(\A) $.
For ${\bf A}^*_\epsilon \Omega_\beta $ we find
\# { {\bf A}^*_\epsilon \Omega_\beta = {1 \over \sqrt {2 \pi}} \int {\rm d}t
\,  \overline { g_\epsilon } (t)   \, {\rm e}^{i {\bf H}_\beta t} {\bf A}^* \Omega_\beta
= \tilde g_\epsilon (-{\bf H}_\beta) {\bf A}^* \Omega_\beta  ,} 
and $\lim_{\epsilon \to 0} \| \tilde g_\epsilon (-{\bf H}_\beta) {\bf A}^* \Omega_\beta
- {\bf P}^- {\bf A}^* \Omega_\beta \| = 0.$
We can now exploit the KMS condition so as to obtain
\& {   f_{a, b} (z) 
&=  \lim_{\epsilon \to 0} \Bigl(
(\Omega_\beta
\, , \, {\bf B} {\rm e}^{ i z {\bf H}_\beta}   {\bf A}_\epsilon 
\Omega_\beta)
-
({\rm e}^{ i \bar z {\bf H}_\beta}  {\bf A}_\epsilon^* \Omega_\beta
\, , \, {\bf B}   
\Omega_\beta) \Bigr)
\cr
&=  \lim_{\epsilon \to 0} 
(\Omega_\beta
\, , \, {\bf B} (\1 - {\rm e}^{- \beta {\bf H}_\beta} ) 
{\rm e}^{ iz {\bf H}_\beta} {\bf A}_\epsilon 
\Omega_\beta)
\cr
&=  
(\Omega_\beta \, , \, {\bf B} (\1 - {\rm e}^{- \beta {\bf H}_\beta} ) 
{\rm e}^{ iz {\bf H}_\beta} {\bf P}^{+} {\bf A} \Omega_\beta) ,}
for $\Im z > 0$.  
A similar argument holds for $\Im z < 0$, thus
\# {
f_{a,b} (z) = 
\left\{
\eqalign{
&  {  (\Omega_\beta \, , \, {\bf B} (\1 - {\rm e}^{- \beta {\bf H}_\beta } ) 
{\rm e}^{iz{\bf H}_\beta} {\bf P}^{+} {\bf A} \Omega_\beta) }
\cr
&
 { (\Omega_\beta \, , \, {\bf B} ( {\rm e}^{- \beta {\bf H}_\beta} - \1 ) 
{\rm e}^{iz {\bf H}_\beta} {\bf P}^{-} {\bf A} \Omega_\beta)    } 
}
\right\} 
{\rm \ for \ }  
\left\{
\eqalign{
& { \qquad  \Im z > 0,}  \cr
& { \left\{ \eqalign{ & { \Im z = 0, | \Re z | < \delta ,}  \cr
                      & {  \Im z < 0 .}                                  } \right\} } 
} \right\} 
}
The function $f_{a,b}$ is closely related to the original function $\omega_\beta (b\tau_t(a))$.
The following result expresses $\omega_\beta (b\tau_t(a))$ for $|t| < \delta $ as a finite 
sum of values of the
function $f_{a,b}(z)$ evaluated at~$z = t + i l \beta$ with $l \in \Z \, $.

\Lm{Let $\omega_\beta$ denote a $(\tau, \beta)$-KMS state, 
and let $\Omega_\beta$ be the unique --- up to a phase
--- time-invariant vector in~$\H_\beta$. 
Then, for $a \in \A (\O_1)$, $ b \in \A (\O_2)$ and $|t| < \delta$,  
\& {  \omega_\beta (b \tau_t(a)) - \omega_\beta (b) \omega_\beta (a) &=
\sum_{ l= 0}^{n-1} f_{a,b} (t + i l \beta)  
+ 
\sum_{ l= 1}^{n} f_{a,b} (t - i l \beta)  
\cr
& \qquad +
(\Omega_\beta , \pi_\beta (b) {\rm e}^{- n \beta {\bf H}_\beta} 
{\bf P}^+ \pi_\beta (\tau_t(a)) \Omega_\beta )
\cr
& \qquad +
(\Omega_\beta , \pi_\beta (b) {\rm e}^{ n \beta {\bf H}_\beta} 
{\bf P}^- \pi_\beta (\tau_t(a)) \Omega_\beta )
,}
holds for all $n \in \N$. }
 
\Pr{Set $\pi_\beta (a) = {\bf A}$, 
$\pi_\beta (b) = {\bf B}$.
We decompose $\1 $ into ${\bf P}^+$, ${\bf P}^-$, and 
${\bf P}_\beta =  | \Omega_\beta ) ( \Omega_\beta |$,
\# {\omega_\beta (b \tau_t(a)) = \omega_\beta (b) \omega_\beta (a) + 
(\Omega_\beta , {\bf B} {\bf P}^+ {\rm e}^{ it {\bf H}_\beta} {\bf A} \Omega_\beta ) +
(\Omega_\beta , {\bf B} {\bf P}^- {\rm e}^{ it {\bf H}_\beta}{\bf A} \Omega_\beta ).}
Lemma 2.1 follows by iteration  of the identities
\& {  (\Omega_\beta, {\bf B} {\bf P}^+ {\rm e}^{ it {\bf H}_\beta} {\bf A} \Omega_\beta) & =
(\Omega_\beta, {\bf B} (\1 - 
{\rm e}^{-  \beta {\bf H}_\beta } ) {\bf P}^+ {\rm e}^{ it {\bf H}_\beta} {\bf A} \Omega_\beta)
\cr
& \qquad +
(\Omega_\beta, {\bf B}  
{\rm e}^{-  \beta {\bf H}_\beta } {\bf P}^+ {\rm e}^{ it {\bf H}_\beta} {\bf A} \Omega_\beta), }
and
\& { (\Omega_\beta, {\bf B} {\bf P}^- {\rm e}^{ it {\bf H}_\beta} {\bf A} \Omega_\beta) & =
(\Omega_\beta, {\bf B} ({\rm e}^{ - \beta {\bf H}_\beta } - \1 )  
{\rm e}^{ (it + \beta) {\bf H}_\beta } {\bf P}^- {\bf A} \Omega_\beta)
\cr
& \qquad +
(\Omega_\beta, {\bf B}
{\rm e}^{ (it + \beta) {\bf H}_\beta} {\bf P}^-  {\bf A} \Omega_\beta)  } 
from relation (25). Both identities hold for $|t| < \delta$.}

Bounds for $|f_{a, b} (i l \beta)|$,  $ l \in \Z \, $ will be derived from

\Lm{Let ${\cal Q} =
\{ z \in \C | \, | \Re z | < \delta, | \Im z | < \delta \}$ be an open square centered
at the origin. Let
$f \colon {\cal Q} \to \C$ be a bounded function 
analytic ${\cal Q}$ and
assume that there exists numbers $C_1 >0 $ and $m > 0$ such that 
for $\Im z \ne 0$, we have
\# {|f(z)| \le C_1  | \Im z | ^{-m}  . }
Let $r \in \R$, $|r| \le \delta$. Then
\# { | f( i r) | \le C_1 \Bigl({\delta \over \sqrt 5} \Bigr)^{-m }  .}
}

\Pr{The function 
\# {g(z) = f(z) (\delta + z)^m (\delta - z)^m }
is analytic in the open square ${\cal Q}$ and continuous at the boundary
of the  square $\partial {\cal Q}$.
Hence assumption  (30) implies bounds for $g(z)$ at the boundary of the square:
\# { | g(z) | \le  C_1  ( \sqrt {5}  \delta)^m , \qquad z \in \partial {\cal Q}.}
By the maximum modulus principle these bounds also hold inside the square. 
Finally $f(ir) = g(ir)   (\delta^2 + r^2)^{-m} $ for $r \in \R \cap {\cal Q}$ and (31) follows
from (33).}

\Hl{Decay of Spatial Correlations}
 
\noindent
In the last section we have seen that locality together with the KMS condition  
implies that the function
\# { t \to \omega_\beta (b \tau_t (a)) - \omega_\beta (b)\omega_\beta (a) }
can be analytically continued into the infinitely often cut plane  
\# { {\cal I}_\delta = \C \backslash \{ z \in \C | 
\Im z = k \beta, k \in \Z, |\Re z| \ge \delta \} .} 
For $|t| < \delta$ the function (34) can be written as an infinite  sum,
\& {  \omega_\beta (b\tau_t(a)) - \omega_\beta (b) \omega_\beta (a) &=
\sum_{ l= - \infty}^{\infty} f_{a,b} ( t + i l \beta) ,  }
involving the function $f_{a,b} $, 
which is analytic on the twofold cut plane
\# { {\cal P}_\delta = \C \backslash \{ z \in \C | \Im z = 0,  |\Re z | \ge \delta \} .}
In fact, Lemma 2.1 even provides an explicit expression for the remainder, if we
prefer to deal with a finite sum:
\& { \Bigl| \omega_\beta (ba) - \omega_\beta (b) \omega_\beta (a) \Bigr|
&\le
\sum_{ l= -n}^{n-1} | f_{a,b} (i l \beta) | +
\| {\rm e}^{- {n \beta \over 2} {\bf H}_\beta} {\bf P}^+ 
\pi_\beta (b^*) \Omega_\beta   \|
 \, 
\| {\rm e}^{- { n \beta \over 2} {\bf H}_\beta} {\bf P}^+ 
\pi_\beta (a) \Omega_\beta   \|
\cr
& \qquad +
\| {\rm e}^{ {n \beta \over 2} {\bf H}_\beta} {\bf P}^- 
\pi_\beta (b^*) \Omega_\beta   \|
 \,  \| {\rm e}^{ {n \beta \over 2} {\bf H}_\beta} {\bf P}^- 
\pi_\beta (a) \Omega_\beta \| ,}
for all $n \in \N$. Bounds for $| f_{a,b} (i l \beta) | $, $l \in \Z$, will be derived
from information on the energy level density of the local
excitations of a KMS state. As expected, the decay of spatial correlations depends on the
infrared properties of the model and 
the essential ingredients for the following theorem are the continuity properties of 
the spectrum of ${\bf H}_\beta$  near zero.

For most applications it is sufficient to consider the following geometric situation:
Let $\O  \subset \R^4$  denote an open and bounded space-time region. 
Consider two space-like separated space time regions $\O_1$, $\O_2$, which can be embedded into
$\O$ by translation; i.e., 
$\O_1 + t e \subset \O_2'$ for all $|t| \le \delta$, and 
$\O_i + x_i \subset \O$, $i=1,2$, for some $x_i \in \R^4$. Under these 
geometric assumptions the following result holds:

\Th{Let $\Omega_\beta$ denote the unique --- up to a phase ---
normalized eigenvector with eigenvalue $\{ 0 \} $ of ${\bf H}_\beta$ in 
the GNS representation $(\H_\beta, \pi_\beta, \Omega_\beta)$ associated
with a $(\tau, \beta)$-KMS state. 
If there exist  positive constants $m>0$ and $C_1 ({\cal O})$
such that   
\# { \|  {\rm e}^{- {\lambda \over 2} |{\bf H}_\beta|} (\pi_\beta (a) - \omega_\beta (a) \1 ) 
\Omega_\beta  \|  
\le C_1({\cal O})  \lambda^{-m}  \, \| a \|, \quad \forall a \in \A (\O), }
holds, then the correlations  
of two local observables $a \in \A (\O_1)$, $b \in \A (\O_2)$, --- $\O_1$, $\O_2$
as described above ---  
decrease like
\# { |  \omega_\beta (ba) - \omega_\beta (b) \omega_\beta (a) | \le  
C_2 (\beta, \O )  (k \beta)^{- 2m}
 \| a \| \, \| b \| , }
for $\delta > \beta$ with   
$ k  < \delta / \beta \le (k+1) $, $k \in \N$.
The constant $C_2(\beta, \O ) \in \R^+$ is independent of $\delta$, $a$ and $b$.}

\Rem{For $\delta$ sufficiently large  compared to the 
inverse temperature $\beta= 1/ k_B T$, the correlations decrease
like $\delta^{- 2m}$. For $\delta < \beta$, we find
\# { \Bigl| \omega_\beta (ba) - \omega_\beta (b) \omega_\beta (a) \Bigr|
\le  4 C( \O , \beta ) \Bigl({\delta \over \sqrt 5} \Bigr)^{-(2m + 1) }
  \| a \| \, \| b \| + \epsilon  \| a \| \, \| b \|, }
where $\epsilon = 2(C_1 ({\cal O}))^2 \beta^{-2m} $.
This fact indicates that one has better decay properties of the correlations in the limit
of zero temperature, $\beta^{-1} \to 0$.
The constant $C_1({\cal O})$ may, as indicated, 
depend on the size of
$\O$, but not on the particular choice of $a, b \in \A (\O)$. The constant $m>0$ may depend on
the size of~$\O$, but we expect that $m$ becomes independent of the size of~$\O$ for
$\O$ sufficiently large. Both $m$ and 
$C_1({\cal O})$ will in general depend on $\beta$.}

\Pr{We proceed in several steps.
\vskip .2cm
\noindent 
(i) By definition ${\bf P}^\pm$ projects 
onto the {\sl strictly} positive resp.\ negative spectrum of ${\bf H}_\beta$. Thus
${\bf P}^\pm \Omega_\beta = 0$, and the thermal
expectation values of the local observables can be subtracted; for instance
\& { \| {\rm e}^{\mp { \lambda \over 2} {\bf H}_\beta} {\bf P^\pm} \pi_\beta (a) 
\Omega_\beta \| 
&
= \| {\rm e}^{\mp { \lambda \over 2} {\bf H}_\beta } {\bf P^\pm}  
\bigl( \pi_\beta (a) - \omega_\beta (a) \1 \bigr)
\Omega_\beta \| 
\cr
& \le  \| {\bf P}^\pm \|  
\| {\rm e}^{- {\lambda \over 2} |{\bf H}_\beta |}   \bigl( \pi_\beta (a) - \omega_\beta (a) \1
\bigr)
\Omega_\beta \| 
\cr
& \le C_1 ({\cal O})  \lambda^{-m}  \, \| a \|, 
\qquad \forall a \in \A(\O) . }
This provides a bound for the remainder in (38), namely
\# { \Bigl| \omega_\beta (ba) - \omega_\beta (b) \omega_\beta (a) \Bigr|
\le
\sum_{ l= -n}^{n-1} | f_{a,b} (i l \beta) | 
+ 2  \bigl( C_1 ({\cal O})  (n \beta )^{-m} \bigr)^2  \| a \| \, \| b \| .}
Note that the number $2n$ of terms 
in this sum has not been fixed yet.
\vskip .2cm
\noindent
(ii) 
The spectral properties of ${\bf H}_\beta$ imply bounds for 
$\sup_{t \in \r}|f_{a,b} (t + i \lambda) |$,
namely  
\& {  \sup_{t \in \r} |f_{a,b} (t + i \lambda) |
& = \sup_{t \in \r}
\Bigl|  \bigl( {\rm e}^{- { \lambda \over 2} {\bf H}_\beta } {\bf P}^{+} 
\pi_\beta (b^*)  
\Omega_\beta \, , \, 
{\rm e}^{it {\bf H}_\beta } 
(\1 - {\rm e}^{- \beta {\bf H}_\beta } )  
{\rm e}^{- { \lambda \over 2} {\bf H}_\beta } {\bf P}^{+} 
  \pi_\beta (a) \Omega_\beta \bigr) \Bigr|
\cr
&
\le  \| {\rm e}^{- { \lambda \over 2}{\bf H}_\beta   } (\1 - {\rm e}^{- \beta {\bf H}_\beta  } )
{\bf P}^{+}  \pi_\beta (a) \Omega_\beta  \| \, 
\| {\rm e}^{- { \lambda \over 2}  {\bf H}_\beta   }
{\bf P}^{+} \pi_\beta (b^*)  \Omega_\beta  \| 
\cr
&\le C (\O , \beta )   \lambda^{-(2m + 1)}  \| a \| \, \| b \| .} 
This can be seen as follows.
Set $\Psi =\pi_\beta (a) \Omega_\beta$. The spectral representation of ${\bf H}_\beta$
yields
\& { \| {\rm e}^{- { \lambda \over 2}{\bf H}_\beta   } (\1 - {\rm e}^{- \beta {\bf H}_\beta  } )
{\bf P}^{+}  \pi_\beta (a) \Omega_\beta  \|^2
&= \int_0^\infty {\rm d} \mu_\Psi  (\nu) \, \,
{\rm e}^{- { \lambda }\nu   } (1 - {\rm e}^{- \beta \nu  } )^2
\cr
&\le \sup_{\nu'} {\rm e}^{- \lambda \nu' /2  } (1 - {\rm e}^{- \beta \nu'  } )^2
\int_0^\infty {\rm d} \mu_\Psi  (\nu) \, \, {\rm e}^{- \lambda \nu /2  } 
\cr
&\le c_\beta \lambda^{-2}
\| {\rm e}^{- { \lambda \over 4}{\bf H}_\beta   } 
{\bf P}^{+}  \pi_\beta (a) \Omega_\beta  \|^2
\cr
&  \le \bigl( C (\O , \beta )   \lambda^{- (m + 1) }  \| a \| \bigr)^2 .}
Bounds similar to (44) hold for $\sup_{t \in \r} |f_{a,b} (t - i \lambda) |$. 
(iii) Refined bounds on $| f_{a,b} (i r) | $, $r \in \R$, $|r| \le \delta $, 
follow from Lemma 2.2:
\& { | f_{a,b}( i r) | & \le C (\O , \beta ) 5^{m + 1/2}  \delta^{-(2m+1)} 
 \| a \| \, \| b \|, 
\qquad \forall 0 \le r \le \delta.}
Given $\O$, $\beta$, and $m$, the constants $C_1 ({\cal O})$ and
$C( \O , \beta )$ are fixed and
\# {  \Bigl| \omega_\beta \bigl( (b - \omega_\beta (b) ) a \bigr) \Bigr| 
\le  
\inf_{n \beta \le \delta} 2  \| a \| \, \| b \| \Bigl( 2n   
 C( {\cal O} , \beta )  
 \Bigl( {\delta \over \sqrt 5} \Bigr)^{-(2m + 1) } 
+ \bigl( C_1 ({\cal O}) 
 (n \beta )^{-m} \bigr)^2 \Bigr)  . }
For $\delta > \beta$ we set $k = \{ n \in \N | n < \delta / \beta \le n+1 \}$.  
Thus 
\& { \Bigl| \omega_\beta (ba) - \omega_\beta (b) \omega_\beta (a) \Bigr|
& \le
C_2 ( \O , \beta ) 
( k \beta)^{- 2m } 
 \| a \| \, \| b \|.}
The choice $n=1$ in (47) reproduce (41).}

\Rem{As mentioned in the introduction, the correlations for free massless bosons 
in a KMS state
decay only like $\delta^{-1}$. From explicit calculations
one expects that\footnote{$^\dagger$}
{\sevenrm The author thanks D.\ Buchholz for this information.}
\# { \| {\rm e}^{- {\lambda \over 2} |{\bf H}_\beta | }  ( \pi_\beta (a) - \omega_\beta (a) \1 ) 
\Omega_\beta \|
\le {\rm const} \cdot \lambda^{-1/2} \| a \| , \qquad a \in \A (\O).}
Thus Theorem 3.1 gives the optimal exponent $2m = 1$ for $\delta$ large (compared to $\beta$).}

\vskip 1cm

\Hl{Uniform Decay of Spatial Correlations}

\noindent
We start this section with a few remarks on the vacuum sector;
we refer to [BW], [BD'AL], [H] for further discussion and motivation.
If the model under consideration has decent vacuum phase space properties, these  
properties manifest [BW] themselves in the nuclearity of the
map  
\# { \matrix {\Theta^{vac}_{\lambda , {\cal O}} \colon  & \pi(\A(\O))'' & \to & \H \hfill
\cr
& {\bf A}  & \mapsto
& {\rm e}^{- \lambda   {\bf H} } ({\bf A} - (\Omega \, , \, {\bf A} \Omega) \1 ) \, \Omega ,} } 
for $\lambda > 0$. Here $\Omega$ denotes the vacuum vector.
If the nuclear norm obeys
\# { \|  \Theta^{vac}_{\lambda, {\cal O} } \|_1 
\le  {\rm const} \cdot \lambda^{- m } , }
Lemma 3.1 of [D'ADFL] applies and  
\# { \Bigl| (\Omega \, , \sum_{j=1}^{j_\circ}  {\bf A}_j {\bf B}_j  \Omega)
 - \sum_{j=1}^{j_\circ} (\Omega \, , {\bf A}_j  \Omega)   (\Omega \, , {\bf B}_j  \Omega) \Bigr| 
\le {\rm const}' 
\cdot
\delta^{-(2m +1)} 
 \Bigl\| \sum_{j=1}^{j_\circ} {\bf A}_j {\bf B}_j \Bigr\|,  }
for  any $j_\circ \in \N$ and any two families of operators
${\bf A}_j \in \pi(\A(\O_1))''$ and ${\bf B}_j \in \pi(\A(\O_2))'' $;  
$\O_1$ and  $\O_2$ as described in front of Theorem 3.1.
 
In contrast to the non-relativistic case the local von Neumann algebras 
in the vacuum representation $\pi (\A(\O))''$ are in general not of type I. 
Nevertheless, if the nuclearity condition (51) holds, then (52) implies that
the local algebras 
$\pi (\A (\O))''$ can be embedded into factors of type I without essential loss
of the local structure, i.e., for every pair $\O_1 \subset \subset \O_2$
there exists a type I factor $\M( \O_1, \O_2)$ such that
\# { \pi (\A (\O_1))'' \subset \M ( \O_1, \O_2 ) \subset \pi (\A (\O_2 ))''.}
By $\O_1 \subset \subset \O_2$ we mean  
that the closure of the open bounded set $\O_1$
lies in the interior of $\O_2$.

Uniform bounds on the thermal correlation functions
of a KMS state can now be derived from thermal phase space properties.
We recall from [BD'AL] the discussion of these phase space properties: The state
induced by the vector $\pi_\beta (a) \Omega_\beta$, $a \in \A(\O)$, 
represents a localised excitation of the KMS state.
The energy transferred to~$\omega_{\beta}$ can be restricted by taking time-averages 
\# { {1 \over \sqrt {2 \pi}} \int {\rm d}t \, f(t) \pi_\beta (\tau_t(a)) \Omega_\beta 
= \tilde f({\bf H}_\beta) \pi_\beta (a)  \Omega_\beta ,
\qquad a \in \A(\O),}
with suitable testfunctions~$f(t)$, 
whose Fourier transforms~$\tilde f (\nu)$ decrease exponentially.
The assumption that the theory has decent phase space properties can be cast into 
the condition that the maps $\Theta^{therm }_{\lambda, {\cal O}} \colon \A(\O)  \to \H_\beta$, 
\# {  
a   \to {\rm e}^{- \lambda | {\bf H}_\beta |} (\pi_\beta (a)
 - \omega_\beta (a) \1 ) \, \Omega_\beta ,
\quad \lambda > 0,} 
are nuclear for all open, bounded space-time regions $\O \subset \R^4$.
Quantitative information on the  
decay of the correlations 
can be extracted from the  nuclear norm of~$\Theta^{therm }_{\lambda, {\cal O}}$.

\Th{Let $\Omega_\beta$ be the unique --- up to a phase
--- time-invariant vector in~$\H_\beta$ corresponding to the $(\tau, \beta)$-KMS state 
$\omega_\beta$. Given the same geometric situation as in Theorem 3.1, 
we pick a $j_\circ \in \N$ and consider two families of operators $a_j \in \A(\O_1)$ and
$b_j \in \A( \O_2) $, $j \in \{  1 , \ldots , j_\circ \}$. If the nuclear norm
$\|  \Theta^{therm }_{\lambda, {\cal O}}  \|_1$ is bounded by
\# { \|  \Theta^{therm }_{\lambda, {\cal O}}  \|_1 
\le {\cal  C}_1  ({\cal O})  (2\lambda)^{-m}  ,}
for some ${\cal  C}_1 ({\cal O}), m  > 0 $, then for $\delta > \beta$
\# {\Bigl| \omega_\beta \Bigl(\sum_{j=1}^{j_\circ} a_j b_j \Bigr)
-  \sum_{j=1}^{j_\circ} \omega_\beta (a_j) \omega_\beta (b_j)
\Bigr| \le  (k \beta)^{- 2m} 
{\cal  C}_2 ({\cal O}, \beta )   
 \Bigl\| \sum_{j=1}^{j_\circ} a_j b_j \Bigr\|     , }
where $k  < \delta / \beta \le (k+1) $, $k \in \N$. ${\cal  C}_2 ({\cal O}, \beta ) $
does not depend on $\delta$ or $j_\circ$.}

\Pr{The uniform bounds are based on the algebraic independence [Ro] of 
the operators $a_j \in \A(\O_1) $ and
$b_j \in \A(\O_2) $. It implies  
\# { \Bigl\| \sum_{j=1}^{j_\circ}  \phi (a_j)  \psi (b_j) \Bigr\|
\le
\| \phi \|
\| \psi \|
\cdot
\Bigl\| \sum_{j=1}^{j_\circ} a_j b_j \Bigr\|,}
for arbitrary linear functionals $\phi $, $\psi$.
Note that $\A(\O_1) $ and
$\A(\O_2) $ still denote the $C^*$-algebras, and not their weak closures
in the representation $\pi_\beta$. Lemma 2.1 generalizes to
\& { \Bigl| \omega_\beta   \Bigl( \sum_{j=1}^{j_\circ} \bigl( b_j 
- \omega_\beta (b_j) \bigr) a_j \Bigr) 
\Bigr| & \le \Bigl| \sum_{j=1}^{j_\circ}
\bigl(\Omega_\beta , \pi_\beta (b_j)  \bigl( {\rm e}^{- n \beta {\bf H}_\beta} {\bf P}^+ 
+ {\rm e}^{ n \beta {\bf H}_\beta} {\bf P}^-  \bigr) \pi_\beta (a_j) \Omega_\beta \bigr) \Bigr|
\cr
&  \qquad +
\sum_{ l= -n }^{n-1} \Bigl| \sum_{j=1}^{j_\circ}  f_{a_j,b_j} (i l \beta) \Bigr| .}
Introducing sequences of vectors 
$\Phi_i , \Psi_k \in \H_\beta $ and of linear 
functionals $\phi_i \in \pi_\beta (\A(\O_1))^*$,
$\psi_k \in \pi_\beta (\A(\O_2))^*$
corresponding [P] to the nuclear maps $\Theta_1 := \Theta^{therm}_{n \beta /2, {\cal O}_1}$ 
resp.\ $\Theta_2 := \Theta^{therm}_{n \beta /2, {\cal O}_2} \, $,  we find
for the first term in the sum on the r.h.s.\ of equation (59)
\& { \Biggl| \sum_{j=1}^{j_\circ}
 (\Theta_2 (b^*_j)  \, , \,     
{\bf P}^{+} \Theta_1 (a_j) ) \Biggr| 
&= \Biggl| \sum_{i=1}^\infty \sum_{k=1}^\infty \sum_{j=1}^{j_\circ}
  \overline { \psi_i (b^*_j) }  \phi_k (a_j)   ( \Psi_i  \, , \,     
{\bf P}^{+} \Phi_k)    \Biggr| 
\cr
&
\le   \sum_{i} \| \phi_i  \|   \| \Phi_i \|    
\sum_{k} \| \psi_k  \|  \| \Psi_k \|   
\, \Bigl\| \sum_{j=1}^{j_\circ} a_j b_j \Bigr\|
}
Taking the infimum over all suitable sequences of vectors 
and linear functionals we obtain
\# { \Biggl| \sum_{j=1}^{j_\circ}
 (\Theta_2 (b^*_j)  \, , \,     
{\bf P}^{+} \Theta_1 (a_j) ) \Biggr| 
\le  \bigl( {\cal  C}_1 ({\cal O})    ( n \beta)^{-m} \bigr)^2 
\Bigl\| \sum_{j=1}^{j_\circ} a_j b_j \Bigr\| .}
A similar bound holds for the term containing ${\bf P}^{-}$ in (59).
We thus arrive at 
\# { \Bigl| \omega_\beta  \Bigl(\sum_{j=1}^{j_\circ} \bigl( b_j - \omega_\beta (b_j)\bigr)  
a_j \Bigr) \Bigr| 
\le
\sum_{ l= -n }^{n-1} \Bigl| \sum_{j=1}^{j_\circ}  f_{a_j,b_j} (i l \beta) \Bigr| 
+
2  \bigl( {\cal  C}_1 ({\cal O})   ( n \beta)^{-m} \bigr)^2 
\Bigl\| \sum_{j=1}^{j_\circ} a_j b_j \Bigr\| . }
Introducing sequences of vectors 
$\hat \Phi_i, \hat \Psi_k \in \H_\beta$ and of linear 
functionals $\hat \phi_i \in \pi_\beta (\A(\O_1))^*$,
$\hat \psi_k \in \pi_\beta (\A(\O_2))^*$
corresponding to the nuclear maps $\Theta_3 := \Theta^{therm}_{\lambda / 4 , {\cal O}_1}$  resp.\ 
$\Theta_4 := \Theta^{therm}_{\lambda / 2 , {\cal O}_2} $,  we find
\& {\sup_{t \in \r}  \Bigl| \sum_{j=1}^{j_\circ} f_{a_j , b_j} & (t+ i \lambda) \Bigr|
=  \sup_{t \in \r} \Bigl| \sum_{j=1}^{j_\circ}
\bigl( \Theta_4 (b^*_j)  \, , \,  
{\rm e}^{it {\bf H}_\beta }
(\1 - {\rm e}^{- \beta | {\bf H}_\beta | })
{\rm e}^{- {\lambda \over 4} | {\bf H}_\beta | }
{\bf P}^{+} \Theta_3 (a_j) \bigr) \Bigr| 
\cr
&
\le \sup_{t \in \r} \sum_{i,k} \Bigl| 
\sum_{j=1}^{j_\circ}  \overline{ \hat \psi_k(b^*_j) } \hat \phi_i(a_j)  
\bigl( \hat   \Psi_k \, , \,  {\rm e}^{it {\bf H}_\beta } 
(\1 - {\rm e}^{- \beta | {\bf H}_\beta | })
{\rm e}^{- {\lambda \over 4} | {\bf H}_\beta | }
{\bf P}^{+}\hat \Phi_i \bigr) \Bigr|
\cr
&
\le   \|  ( \1 - {\rm e}^{- \beta | {\bf H}_\beta | } )
{\rm e}^{- {\lambda \over 4} | {\bf H}_\beta | } \|
 \sum_{i} \| \hat \phi_i  \|   \| \hat \Phi_i \|    
\sum_{k} \| \hat   \psi_k  \|  \| \hat \Psi_k \|  
\Bigl\| \sum_{j=1}^{j_\circ} a_j b_j \Bigr\|
}
Taking the infimum over all suitable sequences of vectors 
and linear functionals we obtain
\# {\sup_{t \in \r}  \Bigl| \sum_{j=1}^{j_\circ} f_{a_j , b_j}  (t+ i \lambda) \Bigr|
 \le   {\cal  C} ({\cal O}, \beta )  
  \lambda^{- 1 }
  \Bigl( {\lambda \over 2} \Bigr)^{- m }
  \lambda^{- m }
\Bigl\| \sum_{j=1}^{j_\circ} a_j b_j \Bigr\| ,}
and similarly $\sup_{t \in \r} \Bigl| \sum_{j=1}^{j_\circ} f_{a_j, b_j} (t - i \lambda) \Bigr|
\le 2^m {\cal  C}  ({\cal O}, \beta )   \lambda^{-(2m+1)} 
\bigl\| \sum_{j=1}^{j_\circ} a_j b_j \bigr\| $.
Theorem 4.1 has now been reduced to the situation discussed in the proof of Theorem 3.1.}  

\Rem{The constant $m> 0$ depends on  
the dynamics of the model under consideration.
Up to now $m$ has not been computed in a thermal state for any model. For the
vacuum, Buchholz and Jacobi [BJ] computed for
free massless bosons  $m= 3$, for $r \ge \lambda$, where $r$ denotes the diameter of $\O$, 
and $m= 1$, for $r < \lambda$.
For the free electromagnetic field they found
$m= 3$, for $r \ge \lambda$, and $m= 1$, for $r < \lambda$. 
Note that in these computations the vacuum expectation values were not subtracted in the
definition of the maps $\Theta^{vac}_{\lambda, {\cal O}}$. 
}

\noindent
{\it  Acknowledgments.\/}
\noindent
A preliminary version was based on
material contained in the thesis of P.\ Junglas. The present formulation arose from
discussions with D.\ Buchholz, whom I want to thank for several hints and
constructive criticism. I am also grateful for the hospitality
extended to me
by the members of the II.\ Inst.\ f.\ Theoretical Physics, 
Universit\"at Hamburg, during the last two years.  
For critical reading of the manuscript many thanks are due to R.\ Verch and J.\ Yngvason.
This work was financed 
by an ``Erwin Schr\"odinger Fellowship'' granted by 
the Fond zur F\"orderung der Wissenschaftlichen Forschung in Austria, Proj.\ Nr.\ PHY-961.  

\vskip 1cm

\noindent
{\fourteenrm References}
\nobreak
\vskip .3cm
\nobreak
\halign{   &  \vtop { \parindent=0pt \hsize=33em
                            \strut  # \strut} \cr 
\REF
{AHR}
{Araki, H., Hepp, K., Ruelle, D.}  
                          {On the asymptotic behavior of Wightman functions
                           in space-like directions}
                          {Helv.\ Phys.\ Acta}
                          {35}     {164--174}
                          {1962}
\REF
{BB}
{Bros, J.\ and Buchholz, D.}       {Towards a relativistic KMS-condition}
                                  {Nucl.\ Phys.\ B} 
                                  {429} {291-318}
                                  {1994}
\REF
{BD'AL}
{Buchholz, D., D'Antoni, C.\ and Longo, R.}
                                    {Nuclear maps and modular structure II:
                                     Applications to quantum field theory}
                                    {\CMP}
                                    {129}  {115--138} 
                                    {1990}
\BOOK
{BR}  
{Bratteli, O.\ and Robinson, D.\ W.} {Operator Algebras and Quantum Statistical Mechanics~I,II} 
                                  {Sprin\-ger-Verlag, New York-Heidelberg-Berlin} 
                                  {1981}
\REF
{BJ}
{Buchholz, D.\ and Jacobi, P.}   {On the nuclearity condition for massless fields} 
                                  {Lett.\ Math.\ Phys} 
                                  {13} {313--323}
                                  {1987}
\REF
{BW}
{Buchholz, D.\ and Wichmann, E.}   {Causal independence and the energy-level
                                   density of states in local quantum field theory} 
                                  {\CMP} 
                                  {106} {321--344}
                                  {1986}
\REF
{D'ADFL}
{D'Antoni, C., Doplicher, S., Fredenhagen, K.\ and Longo, R.}     
                                   {Convergence of local charges and continuity 
                                    properties of W*-inclusions}
                                   {\CMP}
                                   {110}    {325--348}
                                   {1984}
\BOOK
{E}
{Emch, G.\ G.}           {Algebraic Methods in Statistical Mechanics and Quantum Field Theory}
                        {Wiley, New York}
                        {1972}
\REF
{F}
{Fredenhagen, K.}      {A remark on the cluster theorem}
                       {\CMP}
                       {97} {461-463}
                       {1985}
\BOOK
{H}
{Haag, R.}    {Local Quantum Physics: Fields, Particles, Algebras} 
              {Springer-Verlag, Berlin-Heidelberg-New York} 
              {1992}
\REF
{HHW}
{Haag, R., Hugenholtz, N.\ M.\ and Winnink, M.}
                          {On the equilibrium states in quantum statistical mechanics}  
                          {\CMP}	
                          {5}	{215--236}
                          {1967}
\REF
{HK}
{Haag, R.\ and Kastler, D.}        {An algebraic approach to quantum field theory}
                                   {\JMP}
                                   {5}  {848--861}
                                   {1964}
\REF
{HKTP}
{Haag, R., Kastler, D.\ and Trych-Pohlmeyer, E.\ B.}    {Stability and equilibrium states}
                                                      {\CMP} 
                                                      {38} {173--193}
						                                                {1974}
\BOOK
{J}
{Junglas, P.}   {Thermodynamisches Gleichgewicht und Energiespektrum in der 
                 Quantenfeldtheorie} 
                {Dissertation, Hamburg}
                {1987}
\REF
{K}
{Kubo, R.}    {Statistical mechanical theory of irreversible prozesses I.}    
              {J.\ Math.\ Soc.\ Jap}                                       
              {12} {570--586}                                       
              {1957}                                       
\REF
{MS}
{Martin, P.\ C.\ and Schwinger, J.}   {Theory of many-particle systems. I}
                                    {Phys.\ Rev}
                                    {115/6} {1342--1373}
                                    {1959}
\REF
{N}
{Narnhofer, H.}   {Kommutative Automorphismen und Gleichgewichtszust\"ande}
                  {Act.\ Phys.\ Austriaca}
                  {47}   {1--29} 
                  {1977}
\REF
{O}
{Ojima, I.}   {Lorentz invariance vs. temperature in QFT}
              {Lett.\ Math.\ Phys}
              {11}    {73--80}
              {1986}
\BOOK
{P}
{Pietsch, A.}     {Nuclear locally convex spaces} 
                  {Springer-Verlag, Berlin-Heidelberg-New York}  
                  {1972}
\BOOK
{R}
{Ruelle, D.} {Statistical Mechanics: Rigorous Results}
{Addison-Wesely Publishing Co., Inc.}
{1969, and 1989}
\REF
{Ro}
{Roos, H.}   {Independence of local algebras in quantum field theory}
                                    {\CMP}
                                    {16} {238--246}
                                    {1970}
\BOOK
{S}
{Sewell, G.\ L.}   {Quantum Theory of Collective Phenomena}
                 {Clarendon Press, Oxford} 
                 {1986}
\BOOK
{SW}
{Streater, R.F. and Wightman, A.\ S.}   {PCT, Spin and Statistics and all that}
                 {Benjamin, New York} 
                 {1964}
\BOOK
{T}
{Thirring, W.}       {A Course in Mathematical Physics IV} 
                     {Springer-Verlag, Berlin-Heidel\-berg-New York}  
                     {1983}
\cr}

\bye